\documentclass[conference, 10pt]{IEEEtran}
\IEEEoverridecommandlockouts

\usepackage{amsmath,amssymb,amsfonts}
\usepackage{algorithm}
\usepackage{algorithmic}
\usepackage[T1]{fontenc}

\usepackage{tabularx}
\usepackage{xcolor}
\usepackage{booktabs}
\usepackage{soul}
\usepackage{textcomp}

\usepackage{xspace}
\usepackage{graphicx}
\usepackage{color}
\usepackage{xcolor}
\usepackage[sort,compress]{cite}
\usepackage{subcaption}
\definecolor{accessbl}{cmyk}{1, 0.4, 0.3, 0}
\DeclareCaptionFont{bl}{\color{accessbl}}
\usepackage{epsfig,endnotes,paralist,multirow}
\usepackage{wrapfig}
\usepackage{epsfig}
\usepackage[normalem]{ulem}
\usepackage{url}
\usepackage{soul}
\usepackage{multirow}
\usepackage{array}
\usepackage{bbm}
\usepackage{hyperref}

\def\includeComments{include}
\def\includ{include}
\def\comm[#1]{\ifx\includeComments\includ  \bl \texttt{\textbf{\textit{Note: #1}}} \par \fi}
\def\inlinecomm[#1]{\ifx\includeComments\includ  \textit{Note: #1} \fi}

\newcommand{\blue}[1]{\textcolor{black}{#1}}

\newcommand{\green}[1]{\textcolor{black}{#1}}

\newcommand{\para}[1]{\smallskip \noindent {\bf #1}}
\newcommand{\softpara}[1]{\smallskip \noindent \underline{#1}}
\newcommand{\vsp}{\vspace{0.01in}}

\def\blackbox{\hfill {\vrule height6pt width6pt depth0pt}}

\newcounter{theorem}
\newtheorem{thm}{Theorem}

\newtheorem{corollary}{Corollary}

\newtheorem{defin}{Definition}
\newtheorem{ex}{Example}
\newtheorem{lem}{Lemma}
\newtheorem{ob}{Observation}

\newenvironment{thm-prf}{\vsp \begin{thm} \nopagebreak}{\end{thm}}

\newenvironment{lem-wo-prf-box}{\vsp \begin{lem} \nopagebreak}{\end{lem}}
\newenvironment{lem-prf}{\vsp \begin{lem} \nopagebreak}{\end{lem}}

\newenvironment{cor-prf}{\vsp \begin{corollary} \nopagebreak}{\end{corollary}}

\newcounter{packednmbr}
\newenvironment{verypackedenumerate}{
\begin{list}{\thepackednmbr.}{\usecounter{packednmbr}
\setlength{\itemsep}{0pt}
\setlength{\labelwidth}{8pt}
\setlength{\leftmargin}{12pt}
\setlength{\labelsep}{4pt}
\setlength{\listparindent}{\parindent}
\setlength{\parsep}{0pt}
\setlength{\topsep}{1pt}}}{\end{list}}

\newenvironment{packedenumerate}{
\begin{list}{\thepackednmbr.}{\usecounter{packednmbr}
\setlength{\itemsep}{3pt}
\setlength{\labelwidth}{8pt}
\setlength{\leftmargin}{12pt}
\setlength{\labelsep}{4pt}
\setlength{\listparindent}{\parindent}
\setlength{\parsep}{3pt}
\setlength{\topsep}{3pt}}}{\end{list}}

\newcommand{\id}[1]{\mbox{\em #1\xspace}}

\newcommand{\scyqnet}{\texttt{SCY-QNet}\xspace}

\newcommand{\eps}{\mbox{EP}\xspace}
\newcommand{\epss}{\mbox{EPs}\xspace}
\newcommand{\es}{\mbox{ES}\xspace}

\newcommand{\eat}[1]{}

\let\emph\textit

\newcommand{\php}{\mbox{$p_{ob}$}\xspace}          
\newcommand{\bt}{\mbox{$\tau_{b}$}\xspace}         
\newcommand{\ft}{\mbox{$\tau_{f}$}\xspace}         
\newcommand{\bp}{\mbox{$p_{b}$}\xspace}          
\newcommand{\fp}{\mbox{$p_{f}$}\xspace}          
\newcommand{\gt}{\mbox{$\tau_g$}\xspace}      
\newcommand{\gp}{\mbox{$p_g$}\xspace}       
\newcommand{\ct}{\mbox{$\tau_c$}\xspace}      

\newcommand{\swaplatency}{\ensuremath{L_{\mbox{\scriptsize\it swap}}}}

\newcommand{\get}{\mbox{\tt GEM}\xspace}

\newcommand{\mpfont}{\scriptsize}

\ifx\noeditingmarks\undefined
    \newcommand{\MPworker}[2]{{\color{#1}\vrule\vrule}{\marginpar{\color{#1}\mpfont #2}}}
\else
    \newcommand{\MPworker}[2]{}
\fi

\def\BibTeX{{\rm B\kern-.05em{\sc i\kern-.025em b}\kern-.08em
    T\kern-.1667em\lower.7ex\hbox{E}\kern-.125emX}}

\begin{document}


\title{A Comprehensive Protocol Stack for Quantum Networks with a Global Entanglement Module \vspace{-0.1in}}

\author{
\IEEEauthorblockN{Xiaojie Fan\thanks{Corresponding author: Xiaojie Fan (email: xiffan@cs.stonybrook.edu)}, C.R. Ramakrishnan, Himanshu Gupta}
\IEEEauthorblockA{\textit{Department of Computer Science, Stony Brook University, Stony Brook, NY 11790, USA}}
}

\maketitle
\thispagestyle{plain}
\pagestyle{plain}

\begin{abstract}
The development of large-scale quantum networks requires not only advances in physical-layer technologies but also a comprehensive protocol stack that integrates communication, control, and resource management across all layers. We present the first such protocol stack, which introduces a \emph{Global Entanglement Module (GEM)} that maintains a consistent, network-wide view of entanglement resources through distributed synchronization strategies. By enabling real-time adaptive execution of entanglement distribution plans, GEM bridges the gap between static planning and dynamic operation. The stack naturally supports pre-distributed entanglement, purification, and multipartite state generation, making it applicable to a broad range of quantum networking applications. We design and evaluate multiple adaptive heuristics for real-time execution and show that a lightweight \emph{scoring-based strategy} consistently achieves the best performance, improving entanglement generation rates by about 20\% over a globally optimal but non-adaptive fixed-tree baseline and achieving more than a two-fold improvement relative to recent connectionless approaches. Across all scenarios---including predistribution and fidelity analysis---GEM consistently enables lower latency and robust operation. These results establish a practical pathway toward scalable, adaptive quantum internet systems.


\end{abstract}




\section{\bf Introduction}

Quantum networks promise transformative capabilities, including distributed quantum computing, secure communication beyond classical limits, and novel applications based on multipartite entanglement~\cite{qkd,sundaram2024distributing,bartolucci2023fusion}. Realizing these visions requires not only advances in physical-layer technologies---such as quantum repeaters, photonic interfaces, and quantum memories---but also a protocol architecture that can manage communication, control, and resource allocation across heterogeneous systems. Just as the classical Internet emerged from the development of layered protocol stacks, scalable quantum networks will depend on a comprehensive stack that integrates these diverse functionalities.

Several efforts have proposed preliminary protocol stacks for quantum networks, often inspired by the classical Internet layering model~\cite{conext20,sigcomm19,pompili2022experimental}. While these works provide valuable starting points, they typically focus on high-level abstractions without addressing the distinct requirements of entanglement-based communication. In particular, existing proposals do not separate \emph{planning} (how entanglement should be established) from \emph{execution} (how it must adapt under stochastic generation and decoherence), a distinction that becomes fundamental in repeater-based architectures. Our design introduces this separation and augments it with a \emph{Global Entanglement Module (GEM)}, which maintains a consistent, network-wide view of entanglement resources and actively synchronizes updates across nodes. This cross-layer module enables real-time adaptive execution of distribution plans and supports features absent from prior stacks, including \emph{pre-distributed entanglement}, \emph{purification}, and \emph{multipartite state generation}.

This paper presents a comprehensive protocol stack for quantum networks, unifying communication, control, and entanglement management across layers. The main contributions are:
\begin{itemize}
\item \emph{Stack architecture:} A layered design that cleanly separates planning from execution while supporting adaptive operation.
\item \emph{Global Entanglement Module (GEM):} A novel system module that actively maintains consistency and freshness of entanglement metadata to enable decentralized yet globally coordinated adaptation.
\item \emph{Adaptive strategies:} We develop and evaluate multiple adaptive heuristics for real-time execution, and show that a lightweight \emph{scoring-based strategy} consistently outperforms alternatives, combining practicality with near-optimal performance.
\item \emph{Broader functionality:} Seamless support for pre-distributed entanglement, purification, and multipartite state generation.
\item \emph{Evaluation:} NetSquid simulations showing that our scoring strategy improves entanglement generation rates by $\sim$20\% over a globally optimal but non-adaptive fixed-tree baseline, and more than doubles performance relative to recent connectionless approaches.
\end{itemize}
Unlike prior efforts that address isolated layers or specific protocols, our approach offers a unified framework that spans the entire quantum network stack. This positions our work as a foundation for scalable quantum communications and a roadmap for future system implementations.

The remainder of this paper is organized as follows: \S\ref{sec:background} reviews related work and background; \S\ref{sec:Vision} introduces the proposed protocol stack; \S\ref{sec:Functionality} details entanglement distribution via adaptive execution and the GEM; \S\ref{sec:other-apps} discusses other advanced applications; \S\ref{sec:eval} presents simulation results; and \S\ref{sec:conc} concludes with a discussion and future directions.

\section{\bf Preliminaries}
\label{sec:background}

\begin{figure}
    \centering
    \includegraphics[width=0.5\textwidth]{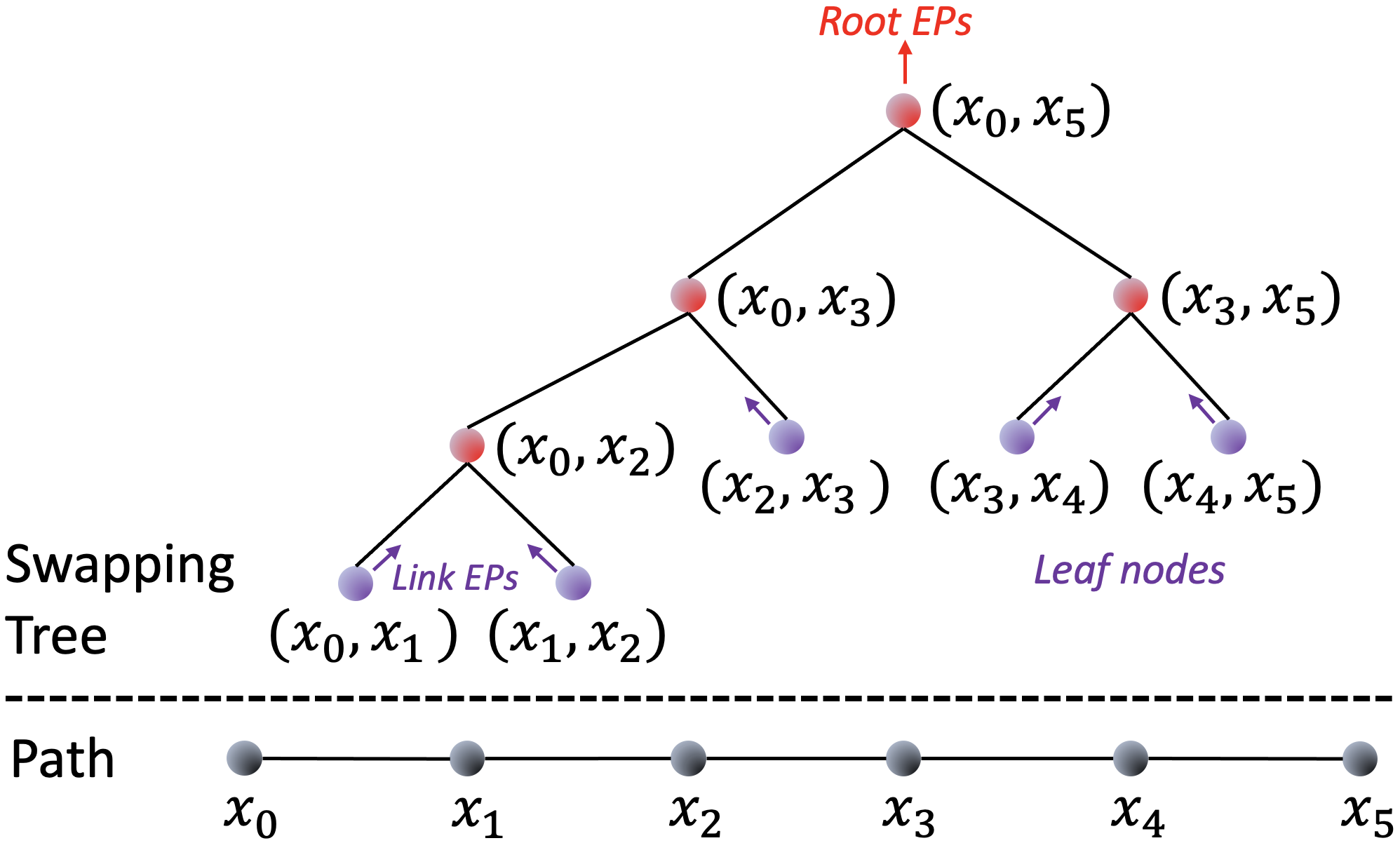}
    \caption{A swapping tree over a path. Binary internal nodes perform entanglement swapping. The leaves of the tree generate link EPs, and internal nodes generate remote EPs.}
    \vspace{-0.2in}
\label{fig:tree}
\end{figure}

\para{Entanglement Generation via Swapping Trees.}
To generate a remote entangled pair (EP), the quantum network performs a series of entanglement swapping operations. This process can be represented as a swapping tree~\cite{swapping-tqe-22}, which captures the structure of intermediate operations and key performance metrics, such as generation latency (see Fig.~\ref{fig:tree}).

\softpara{Swapping Trees.} 
 An efficient way to generate an \eps over a pair of remote network nodes $(s, d)$ using \epss over network links (i.e., edges) is to: 
(i) create a path $P$ in the network graph from $s$ to $d$  with \epss over each of the path's edges, and 
(ii) perform a series of entanglement swapping (ES) over these \epss. 
The \es operations over $P$ can be performed in any order. 
The order can be represented by an 
\emph{entanglement-swapping tree}.  Formally, an entanglement-swapping tree is a complete binary tree consisting of two kinds of nodes:
\begin{itemize}
    \item leaves of the form $\id{link}(x_i,x_j)$ representing link \epss, 
    ($(x_i, x_j) \in E(Q)$ ) and
    \item internal nodes of the form $\id{swap}(x_i, x_j, x_k)$ with $x_i, x_j, x_k \in V(Q)$ representing \epss $(x_i, x_j)$ generated by \es between $(x_i, x_k)$ and  $(x_k, x_j)$ at network node $x_k$. The two children of such a node will be entanglement swapping trees, each rooted at $\id{link}(x_i, x_k)$ or $\id{swap}(x_i, x_k, \_)$, and
    $\id{link}(x_k, x_j)$ or $\id{swap}(x_k, x_j, \_)$.  
\end{itemize}

\softpara{Generation Latency and Generation Rate.}
The latency of generating \epss from an entanglement-swapping tree can be recursively computed as follows (from~\cite{swapping-tqe-22}). 
Let $t$ be an entanglement-swapping tree with a $\id{swap}$ node at its root. Let the two children of $t$ be $l$ and $r$. If the mean generation latencies of $l$ and $r$ are noted by $l_l$ and $l_r$ then the generation latency of $t$ can be estimated as $l_t = \swaplatency(l_l, l_r)$ where
\begin{equation}
\swaplatency(l_l, l_r) = (\tfrac{3}{2} \max(l_l, l_r) + \ft + \ct)/\fp, \label{eqn:EP_swaplatency}
\end{equation}
where 
\ft is the latency of \es,   
\fp is the probability of success of \es, and 
\ct is the latency of classical communication. 
The $\tfrac{3}{2}$ factor comes from the fact that given two exponential distributions with similar means $\mu_1 \approx \mu_2$, the mean of the \emph{maximum} of the two is $\approx \tfrac{3}{2}\max(\mu_1, \mu_2)$.

In the base case $t=\id{link}(x_i, x_j)$, latency $l_t = \tfrac{1}{\id{rate}(x_i, x_j)}$. 

\para{Adaptive Strategies for Entanglement Generation.}
Recent adaptive routing strategies~\cite{real-time,mdp} adopt a two-stage approach that separates offline planning from real-time execution. In the offline stage, an offline plan is computed based on the current network topology and resource estimates. This plan defines both the expected link-level EP generation rates and a default swapping schedule.

During the online/real-time execution stage, the actual availability of entangled pairs may deviate from the offline plan due to stochastic generation outcomes. Adaptive protocols monitor the entanglement state in real time and dynamically adjust swapping decisions to prioritize available or higher-quality EPs. This decoupling enables more efficient entanglement distribution, improving both throughput and responsiveness to changing network conditions.


\para{Purification-Augmented Swapping Trees (PAST).}
\green{Entanglement fidelity quantifies how close a quantum state is to the ideal target state; it degrades in practice due to decoherence and operational noise. To counteract this, purification consumes multiple lower-fidelity copies to probabilistically distill a higher-fidelity state~\cite{gu2024fendi}.}
When purification is required, we extend this structure into what we call a \emph{purification-augmented swapping tree (PAST)}. In a PAST, additional unary nodes are inserted to represent purification operations: each node consumes two or more entangled pairs and, if successful, outputs a higher-fidelity pair for subsequent swapping. This unified representation captures both swapping and purification within a single formalism and provides a convenient framework for analyzing entanglement latency, success probability, and resource usage.

\section{\bf Proposed Quantum Network Stack}
\label{sec:Vision}

\subsection{\bf Motivation and Justification}

Quantum network services fundamentally rely on generating entangled pairs (EPs) between distant nodes. Because entanglement degrades rapidly due to decoherence, latency and fidelity are critical performance factors. Meeting these requirements requires not only physical-layer advances but also a protocol architecture that coordinates efficient generation strategies — such as optimized swapping and purification plans — and supports adaptive resource use.

Our stack design is motivated by the following four key requirements for scalable and robust EP generation:

\begin{enumerate}
\item \textbf{Plan Creation:} Before execution, the network must create an EP generation plan based on current resources. This includes (i) selecting a path between source and destination, (ii) allocating generation rates on each link, and (iii) defining a swapping order at intermediate nodes. These decisions determine EP latency/fidelity, formalized as a swapping tree~\cite{swapping-tqe-22}, or a purification-augmented swapping tree~\cite{qcnc-purification}.

\item \textbf{Adaptive Protocols:} Given the probabilistic nature of entanglement generation, runtime adaptation is crucial. Rather than rigidly following a static plan, protocols may adjust the swapping order in response to real-time conditions, such as EP availability or qubit age~\cite{real-time,mdp}. For example, prioritizing older or \green{fresher} EPs may improve fidelity and throughput. Such adaptive decisions 
necessitate a global view of entanglement resources---motivating the need for a Global Entanglement Module (\get). 

\item \textbf{Pre-Distributed EPs:} Leveraging previously generated EPs can substantially reduce latency during execution~\cite{predist-qce-22}. These pre-distributed EPs act as cached resources along entanglement paths. Their effective use requires protocol-level awareness of where such EPs exist and their fidelity—again necessitating a global view of entanglement resources (\get).

\item \textbf{Multipartite Generalization:} While many communication tasks use bipartite EPs, advanced applications like distributed sensing and measurement-based quantum computing require multipartite states (e.g., GHZ or graph states)~\cite{fan2024optimized}. 
Such states can often be constructed by extending or composing EP-based protocols, reinforcing the need for a flexible and extensible control stack.
\end{enumerate}




\subsection{\bf High-Level Description}

\begin{figure}
    \centering
    \includegraphics[width=0.5\textwidth]{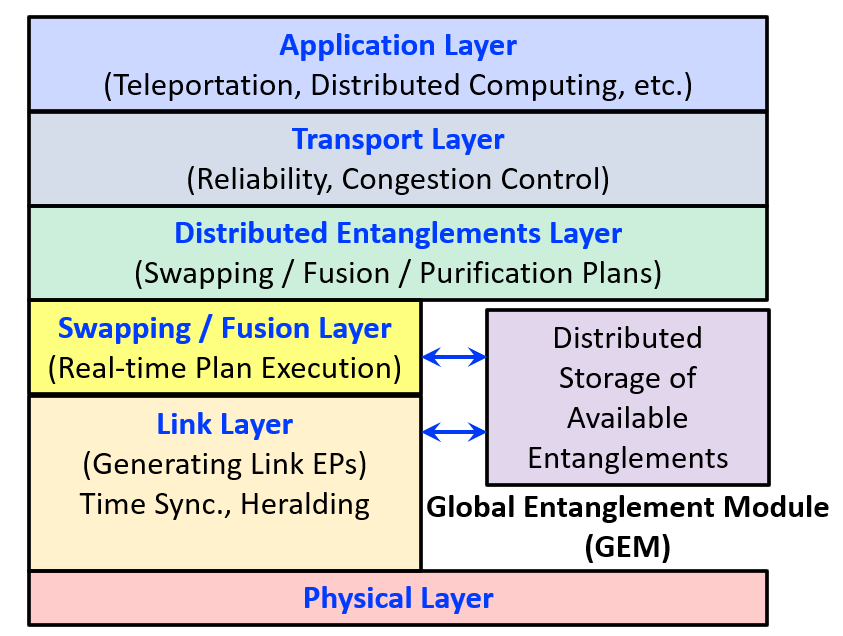}
    \caption{Proposed Quantum Network Stack.}
  \vspace{-0.2in}
  \label{fig:stack_overview}
\end{figure}

We now present a six-layer protocol stack for quantum networks (see Fig.~\ref{fig:stack_overview}). While inspired by the layered design of classical communication systems, this architecture incorporates features unique to quantum entanglement distribution—notably adaptive execution and a cross-layer Global Entanglement Module (\get). 

\begin{verypackedenumerate}
    \item \textbf{Applications Layer:} Provides the interface for quantum applications that request entanglement, passing requests with desired properties (e.g., fidelity, multipartite size) to the lower layers.

    \item \textbf{Transport Layer:} Ensures reliable end-to-end delivery of entanglement and manages basic congestion. It supports applications by abstracting away failures and variability in lower layers.

    \item \textbf{Distributed Entanglements Layer:} Provides entanglement as a network-wide service. For each request, it generates a \emph{plan}, such as an entanglement route~\cite{caleffi}, swapping tree~\cite{swapping-tqe-22}, purification-augmented swapping tree (PAST)~\cite{qcnc-purification}, or fusion tree~\cite{ghz-qce-23}. It also instructs the link layer to create link-level EPs at appropriate rates.\footnote{Although strict layering discourages cross-layer interactions, practical implementations may allow exceptions, e.g., the transport layer issuing instructions directly to the link layer.}


    \item \textbf{Global Entanglement Module (\get):} A cross-layer module that maintains a best-effort, network-wide view of all active entanglements. \get tracks metadata such as states, fidelity, age, and usage, and synchronizes updates across nodes. See~\S\ref{sec:GEM} for details.
    
    \item \textbf{Swapping/Fusion Layer:} Executes the plan in real time by performing entanglement swapping or fusion. This layer is responsible for adapting execution to current network conditions, thereby "operationalizing" the separation of planning (above) and execution (here).

    \item \textbf{Link Layer:} Generates link-level EPs as instructed by higher layers, using tight synchronization of entanglement sources and heralded quantum memories.

    \item \textbf{Physical Layer:} Handles the low-level processes for creating link-level EPs over physical channels.
\end{verypackedenumerate}

\begin{figure}
    \centering
    \includegraphics[width=0.5\textwidth]{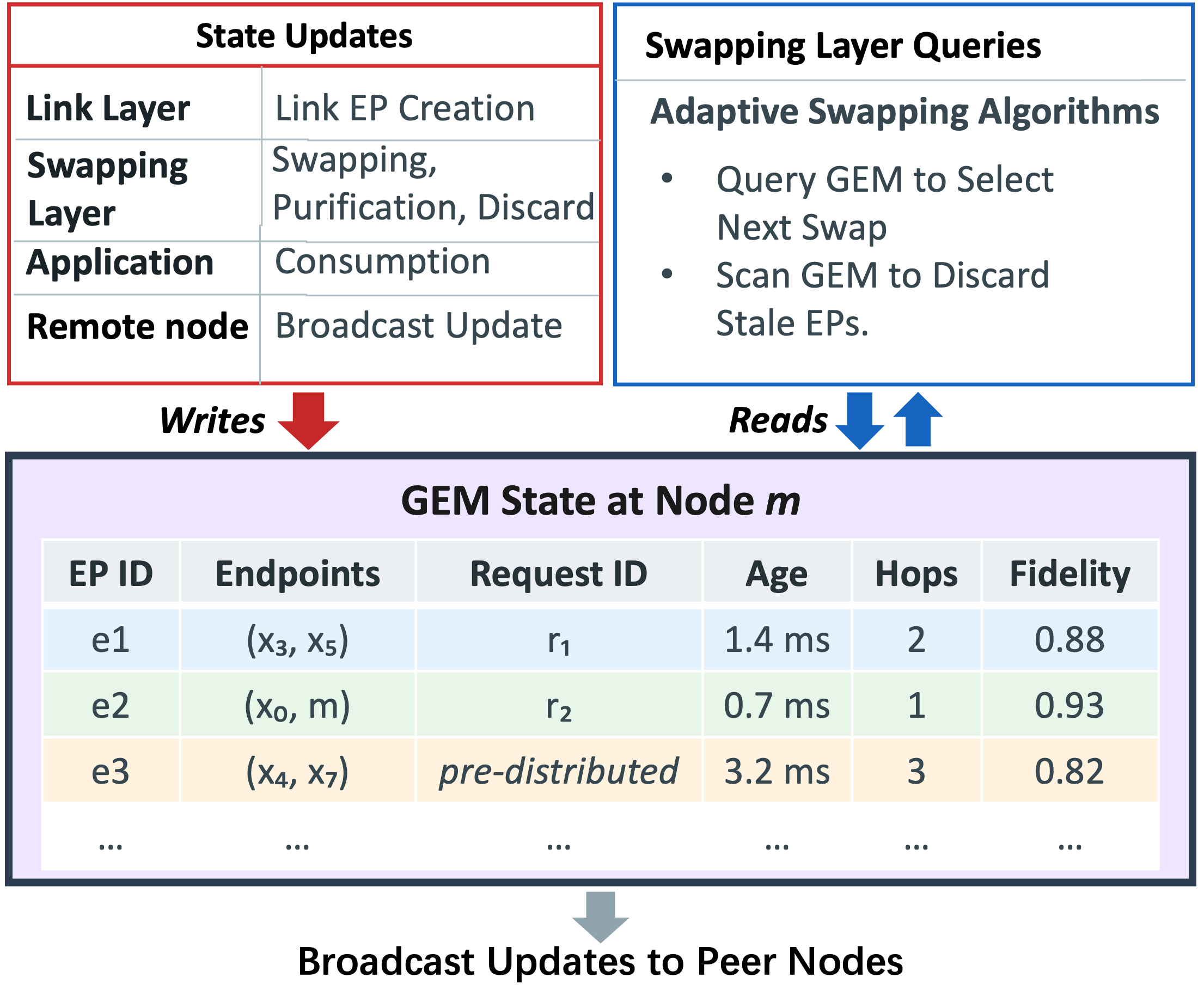}
\vspace{-0.05in}
    \caption{Global Entanglement Module (GEM) at a local node.}
  \vspace{-0.25in}
  \label{fig:gem}
\end{figure}

\subsection{\bf Global Entanglement Module (\get)}
\label{sec:GEM}

The \emph{Global Entanglement Module (\get)} is a cross-layer coordination mechanism 
that maintains a near-real-time, best-effort view of entanglement resources across the network. 
It enables decentralized yet globally consistent decision-making for adaptive execution.

\softpara{Tracked Metadata.}  
Each node maintains a local \get replica that records the state of all active entanglements, including creation time, estimated fidelity, age, endpoints, and the assigned request. This unified view is exposed to upper layers for planning and execution.

\softpara{State Synchronization.}  
Local replicas are synchronized via broadcast updates whenever entanglements are created, consumed, purified, or swapped. Updates carry timestamps to ensure that nodes retain the freshest record, achieving eventual consistency under reliable classical communication. For near-term networks of modest scale (e.g., \scyqnet), broadcast overhead remains tractable.

\softpara{Role in Adaptive Execution.}  
By exposing a shared state abstraction, \get enables adaptive swapping, purification, and predistribution. Critically, \get allows decentralized policies to react to real-time network dynamics without requiring strong global consistency. To our knowledge, no prior architecture has introduced such a unifying module, making \get both novel and essential for scalable quantum networking.
\green{See Fig.~\ref{fig:gem}.}

\begin{table*}[h]
\caption{Comparison of Quantum Network Stack Proposals.}
\label{tab:quantum-stack-overview}
\centering
\renewcommand{\arraystretch}{1.2}
\begin{tabular}{|l|l|l|c|c|c|}
\hline
\textbf{Paper} & \textbf{Architecture} & \textbf{Distribution Strategies} & \textbf{Purification} & \textbf{Multipartite} & \textbf{Pre-Distribution} \\
\hline
Van Meter et al.~(2008)~\cite{van2008system} & Custom Layers & Static Swapping Tree & Yes & No & No \\
Pirker and Dür (2019)~\cite{qn-arch-19} & Custom Layers & Multipartite-Specific Planning & No & Yes & No \\
Dahlberg et al.~(2019)~\cite{sigcomm19} & Link-layer protocol & Swap-ASAP & No & No & No \\
Kozlowski et al.~(2020)~\cite{conext20} & Network-layer protocol & Swap-ASAP & No & No & No \\
Van Meter et al.~(2022)~\cite{qn-arch-22} & SDN-based rulesets & Static Swapping Tree & Yes & No & No \\
Pompili et al.~(2022)~\cite{pompili2022experimental} & Experimental testbed & Swap-ASAP & No & No & No \\
Bacciottini et al.~(2025)~\cite{towsley-24} & Layered TCP/IP-style & Path-Adaptive & No & No & No \\
\textbf{This work} & 
\begin{tabular}[c]{@{}l@{}}\textbf{Layered stack with} \\ \textbf{planning/execution } \\ \textbf{separation via \get} \end{tabular} 
& 
\begin{tabular}[c]{@{}l@{}}\textbf{Static and adaptive plans,} \\ \textbf{purification-augmented, pre-} \\ \textbf{distributed, multipartite} \end{tabular} 
& 
\textbf{Yes} & \textbf{Yes} & \textbf{Yes} \\
\hline
\end{tabular}
\vspace{-0.2in}
\end{table*}
\subsection{\bf Applications To Be Supported}

Our stack abstracts entanglement distribution into a unified service, enabling both foundational and advanced quantum networking applications.

{\em Foundational: Entanglement Generation and Teleportation.}
Entangled pair (EP) generation is the core primitive underlying nearly all quantum network tasks. Our stack supports this functionality across multiple execution modes, including static plans, purification, and runtime-adaptive strategies, with the (possibly purification-augmented) swapping tree (see Section~\ref{sec:background}) as the fundamental abstraction. 
Quantum teleportation, which transfers an arbitrary qubit state using classical communication and a shared EP, is then provided as a basic service layered on top of EP generation.

{\em Advanced: Distributed Quantum Computing and Multipartite State Generation.}
Distributed quantum computing (DQC) connects multiple small- or medium-scale processors to collaboratively execute large-scale quantum algorithms~\cite{sundaram2024distributing}. This places stringent demands on entanglement generation rate and fidelity, both of which are addressed by our protocol stack. 
Beyond bipartite use cases, advanced applications such as distributed sensing and measurement-based quantum computing require multipartite entangled states (e.g., GHZ or graph states). Our stack supports these by generalizing EP distribution protocols to multipartite settings, ensuring broad applicability across diverse quantum networking workloads.

\subsection{\bf Related Work}

Prior work on quantum network architectures can be grouped into (i) stack proposals and (ii) protocol-level designs addressing swapping/routing. While these efforts provide valuable building blocks, none offer an integrated protocol stack like ours, which supports effective adaptive execution and a variety of QN applications.


\para{Stack Proposals.}
Several architectural frameworks for quantum networks have been proposed. 
Van Meter and colleagues outlined recursive layering models~\cite{van2008system,van2011recursive,qn-arch-22,matsuo2019quantum} that extend classical Internet principles into the quantum domain. In particular,~\cite{qn-arch-22,matsuo2019quantum} adopts a software-defined networking paradigm and introduces flexible rule sets to coordinate entanglement generation and swapping; however, it supports only static entanglement-swapping trees.
Dahlberg et al.~\cite{sigcomm19} introduced a link-layer protocol for heralded entanglement generation.
Pirker and Dür~\cite{qn-arch-19} propose a hierarchical stack tailored for the generation of multipartite graph states; while valuable, it is limited to a single generation scheme and
does not generalize to unicast EP generation or entanglement routing.
Finally,~\cite{wehner2018quantum} is a broad vision work which offers 
valuable abstractions and directions, but does not address concrete 
stack-layer coordination or stack features for efficient 
entanglement generation.
These efforts illustrate how layering concepts can be adapted to quantum settings and provide useful abstractions for entanglement management. 
\green{However, these efforts remain either high-level or limited in scope. They do not separate planning (e.g., path and swapping-tree construction) from execution (runtime decisions under stochastic conditions), lack a unifying abstraction such as \get to maintain a shared view of entanglement outcomes and dynamic network state, and rarely support pre-distributed entangled pairs or multipartite distribution at the protocol level, as summarized in Table~\ref{tab:quantum-stack-overview}.}

\para{Relevant Protocol Works.}
Although these works do not propose full-stack architectures, they contribute important protocol-layer mechanisms that inform the design of scalable QN stacks. We group them into four main categories:

\emph{Fixed Swapping Trees.} 
Several works compute optimal fixed execution plans or swapping trees for entanglement generation, but do not support runtime adaptation~\cite{sigcomm20,dlcz}. 
Matsuo et al.~\cite{matsuo2019quantum} proposed a RuleSet-based protocol for link-level entanglement generation and purification using fixed execution plans. 
In our prior work~\cite{swapping-tqe-22}, Ghaderibaneh et al.~developed a dynamic programming approach to compute optimal swapping trees under fidelity and capacity constraints. 
Both approaches are limited to static plans and do not integrate into a full stack.

\emph{Swap-ASAP.} 
Other designs adopt a greedy strategy, performing immediate swapping as soon as two entangled pairs become available at intermediate nodes. 
Kozlowski et al.~\cite{conext20} implemented such a fixed-path strategy with deferred Pauli corrections, while Pompili et al.~\cite{pompili2022experimental} demonstrated it experimentally in a three-node network. 
These designs are simple but lack flexibility and coordination across the network.

\emph{Adaptive Routing Strategies.} 
More recent work has focused on adaptive routing and execution under stochastic conditions. 
In our earlier work~\cite{real-time}, Sundaram and Gupta introduced adaptive algorithms that select routes and swapping operations based on EP availability and decoherence.
I{\~n}esta et al.~\cite{mdp} modeled entanglement distribution as a Markov Decision Process, optimizing swapping and memory cutoff policies, but at prohibitive computational cost. 
\green{
Bacciottini et al.~\cite{towsley-24} adopt hop-by-hop swapping with dynamic next-hop selection. However, their sequential design incurs high latency under stochastic BSM operations, as a failure at any hop discards the entire end-to-end entanglement. Moreover, their approach does not generalize beyond bipartite entanglement.}

\emph{Purification-Aware Plans.} 
In a recent work from our group~\cite{qcnc-purification}, Fan et al.~proposed algorithms that embed purification into swapping and fusion trees to support fidelity-constrained generation of bipartite and multipartite states, introducing a level-based structure to coordinate purification across overlapping trees. 
This line of work addresses fidelity constraints but focuses purely on algorithmic design without proposing a corresponding stack architecture.

\section{\bf Entanglement Distribution via Adaptive Execution}
\label{sec:Functionality}


A central contribution of our stack is its ability to support
\emph{adaptive execution} in entanglement distribution.
While prior designs have largely treated entanglement generation as a
static process driven by precomputed plans, our approach separates
\emph{planning} from \emph{execution}, with the Global Entanglement Module (\get) providing the shared state needed for decentralized decision-making.
This separation allows the stack to accommodate the
probabilistic nature of quantum operations, integrate pre-distributed
resources, and adaptively update swapping or purification decisions based
on the evolving network state.
We present specific mechanisms, such as plan construction, adaptive
swapping policies, and proactive resource management as strong,
practical proposals enabled by our architecture, while emphasizing that the architecture itself is general enough to accommodate alternate or
more elaborate mechanisms as the field evolves. 



\subsection{\bf Application Layer: Request Specification}

The Application Layer generates \textit{Entanglement Distribution Requests (EDRs)} for quantum applications. Each EDR includes a unique identifier ({\tt Request ID}) and requirements as described below.
   
\para{Entanglement Requirements.} 
For bipartite entanglement, i.e., EPs, the request specification 
includes the pair of \green{nodes} over which the remote EP is to be
generated/distributed, along with the performance requirements of the entanglement request. Requirement types can be categorized into:
\begin{itemize}
        \item Generating a fixed number of entanglements.
        \item Generating entanglements at a specified rate with maximum achievable fidelity.
        \item Generating entanglements at the maximum possible rate while maintaining a fidelity threshold.
        \item Generating entanglements at a specified rate and fidelity.
        \item {\em Entanglements for Distributed Circuits.} Generating a set of entanglements (possibly, with varying specifications as above), with a given consumption order. This request involves the network determining the ``batches'' in which the entanglements should be generated to enable their most efficient generation, while considering the specified consumption order. This request type is motivated by the application of distributing quantum circuits, and is referred to as {\em Distributed Circuit Entanglements (DCE)}. 
\end{itemize}
The above is accompanied by the following, as required.
\begin{itemize}
    \item \emph{Fidelity threshold}: Specifies the minimum acceptable fidelity for the generated entangled pairs.
    \item \emph{Rate threshold}: Specifies the minimum generation rate required to satisfy the request.
    \item \emph{Priority (optional)}: Specifies the request’s priority in accessing shared resources.
\end{itemize}
These \textit{Entanglement Distribution Requests (EDRs)} are then passed to the Transport Layer, which determines how to process them efficiently while coordinating with lower layers of the stack.  \green{A request, once admitted, remains active until its termination 
condition is met (e.g., the requested count is complete), 
continuously generating EPs throughout its lifetime.}



\begin{algorithm}
\caption{Execution Monitoring in Transport Layer}
\label{alg:execution-monitor}
\textbf{Input:} logs $\mathcal{L}=\{(t_i,f_i)\}_{i=1}^n$; targets $r^*,f^*$; tolerances $\delta_r,\delta_f$; steps $\eta_r,\eta_f$; params $N_{\min},T_{\min},\Delta T,T_{\mathrm{susp}}$; persistent $c{\gets}0$, $t_{\text{last}}$.\\
\textbf{Output:} tuple $(\texttt{action},\texttt{param})$, where \\
\hspace*{1em}$\texttt{action}\in\{\texttt{null},\texttt{replan},\texttt{adj\_rate},$\\
\hspace*{1em}$\texttt{adj\_fid},\texttt{suspend},\texttt{terminate}\}$.
\begin{algorithmic}[1]
\IF{$n{<}N_{\min}$ \OR $t_n{-}t_1{<}T_{\min}$ \OR $t_{\text{now}}{-}t_{\text{last}}{<}\Delta T$}
    \RETURN $(\texttt{null},\texttt{null})$
\ENDIF
\STATE Compute average rate $r$ and fidelity $f$ over recent window
\IF{$|r{-}r^*|\le\delta_r r^*$ \AND $|f{-}f^*|\le\delta_f f^*$}
    \STATE $c\gets 0$; \textbf{return} $(\texttt{null},\texttt{null})$
\ENDIF
\STATE $c\gets c+1$
\IF{$c=1$ \OR $c=5$} \RETURN $(\texttt{replan},\texttt{null})$
\ELSIF{$c\in\{2,3\}$}
    \STATE $\mathcal{S}\gets\{\,(r,r^*,\delta_r,\eta_r,\texttt{adj\_rate}),$
    \STATE $\phantom{\mathcal{S}\gets\{\,}(f,f^*,\delta_f,\eta_f,\texttt{adj\_fid})\,\}$
    \FOR{$(x,x^*,\delta,\eta,\texttt{act})\in\mathcal{S}$}
        \IF{$x<x^*(1{-}\delta)$} \RETURN $(\texttt{act},+\eta)$
        \ELSIF{$x>x^*(1+\delta)$} \RETURN $(\texttt{act},-\eta)$
        \ENDIF
    \ENDFOR
\ELSIF{$c=4$} \RETURN $(\texttt{suspend},T_{\mathrm{susp}})$
\ELSE \RETURN $(\texttt{terminate},\texttt{null})$
\ENDIF
\end{algorithmic}
\end{algorithm}

\subsection{\bf Transport Layer: Processing and ``Congestion'' Control}

The Transport Layer sits between application requests and lower-level execution, ensuring that entanglement delivery remains reliable and efficient despite the stochastic nature of quantum operations. Its responsibilities fall into three categories: (i) processing complex or meta-requests such as Distributed Circuit Entanglement (DCE) requests, (ii) formulating pre-distribution plans to reduce the latency of future demands, and (iii) monitoring execution to detect performance degradation and trigger corrective actions, akin to congestion control in classical networks.

\para{Processing Complex/Meta Requests.} 
The Transport Layer is responsible for processing complex/meta requests, e.g., for {\em Distributed Circuit Entanglement (DCE)} requests described before, the Transport Layer determines how to partition the required entangled pairs into batches that can be efficiently generated while leveraging the specified consumption order and 
without exceeding decoherence limits.
One potential approach to determine batches for DCE requests is to use a dynamic-programming-based method~\cite{sundaram2024distributing,sundaram2025dynamic} that computes an optimal batch segmentation when the consumption order is total~\cite{sundaram2025dqc}. 
In more general settings where the EP consumption order is partial/unknown, a greedy strategy can be used to select low-latency batches while respecting dependency constraints. 

\para{Determining Pre-Distributed Entanglements.} 
The Transport Layer not only handles application-driven entanglement requests but also formulates plans to proactively generate {\em pre-distributed} EPs~\cite{predist-qce-22} between appropriate node pairs to help reduce the generation latency of future anticipated EPs. These pre-distributed EPs are proactively generated and stored between the selected node pairs, enabling them to be used immediately in subsequent entanglement generation processes. 
The generation of such plans depends on anticipated requests, which may be inferred from historical application activity, and requires knowledge of the network's overall state. As such, this functionality is best handled by the Transport Layer. The format of a generation plan for pre-distributed EPs typically includes selected node pairs, the target number of EPs to maintain between each pair, and optional thresholds for fidelity and age. Although these thresholds were not considered in earlier work~\cite{predist-qce-22}, we regard them as essential components of a robust planning strategy.

One potential approach to determining generation plans for such pre-distributed EPs is to utilize the strategies in~\cite{predist-qce-22}, which include a greedy selection method that incrementally identifies node-pairs (referred to as {\em super-links} in~\cite{predist-qce-22}) along disjoint entanglement paths. In each greedy iteration, candidate super-links are evaluated based on their benefit-to-cost ratio, where the benefit reflects the reduction in expected entanglement latency across all request pairs. 
Another approach to determining plans for pre-distributed EPs is to use a clustering-based strategy that partitions the set of entanglement requests into $k$ groups and selects an optimal super-link for each group. 

\para{Execution Monitoring.} 
The Transport Layer manages the lifecycle and enforces reliability for long-running entanglement requests issued by the Application Layer. While the Application Layer defines performance requirements such as fidelity and generation rate, the Transport Layer is responsible for ensuring that these targets are met reliably over time, based on feedback from lower layers. 
To this end, the Transport Layer incorporates an \emph{Execution Monitor} that continuously tracks runtime statistics such as generation rate and estimated fidelity. It evaluates whether the actual system performance during execution aligns with the original request constraints. These responsibilities of the Transport Layer are akin to mitigating and controlling congestion in classical networks.

\softpara{Corrective Actions.}
If a deviation from performance targets is detected, e.g., fidelity falls below the threshold, the Transport Layer may initiate corrective actions such as:
\begin{itemize}
    \item Requesting the Distributed Entanglements and/or Swapping Layer to re-plan the entanglement generation strategy, such as selecting an alternative entanglement route, swapping tree, and/or adaptive execution strategy.
    \item Adjusting the fidelity or rate requirements of the request and triggering a corresponding update in the entanglement generation plan.
    \item Temporarily suspending the request to avoid wasting resources during unfavorable network conditions.
    \item Terminating the request if the performance cannot be recovered, and notifying the Application Layer.
\end{itemize}
To avoid premature reactions to transient fluctuations, the Execution Monitor only begins correction after a minimum observation window has passed, defined either by the number of samples (\(N_{\min}\)) or a minimum time duration (\(T_{\min}\)). Additionally, a cool-down interval \(\Delta T\) is enforced between successive corrective actions to prevent overly frequent adaptations that could destabilize the system.


\green{If the observed rate or fidelity falls short of the required threshold, 
the system increments a retry counter~$c$ and responds based on its new 
value, where $c=k$ denotes the $k$-th corrective action taken in response 
to the current performance deviation. The first failure ($c=1$) triggers 
a simple re-planning; the next two failures ($c=2, 3$) gradually decrease 
or increase the target performance thresholds as needed. If the issue 
persists, the request is temporarily suspended ($c=4$), followed by one 
final attempt at re-planning ($c=5$). If none of these steps succeed, the 
request is ultimately terminated.}

The system supports bidirectional threshold adjustment: if the observed rate or fidelity deviates from its target by more than $\delta$ (default $30\%$ of the target), the threshold is adjusted by $\eta$ (default $10\%$ of the target).
This dual adjustment mechanism balances performance assurance with resource efficiency as shown in Algorithm~\ref{alg:execution-monitor}. 
In short, our design incorporates a hierarchical corrective strategy that escalates from re-planning to threshold adjustments, temporary suspension, and termination.

\para{Summary.} Thus, the Transport Layer provides clean and abstracted feedback to the Application Layer, hiding the stochastic nature of entanglement generation at lower layers. This abstraction enables applications to interact with the quantum network through a reliable and predictable service interface.

\subsection{\bf Distributed Entanglements Layer: Plan Creation}

The Distributed Entanglements (DE) Layer serves as the planning core of the stack, translating application- or transport-layer requests into executable generation plans. These plans specify how entanglement should be established--through routes, swapping trees, or purification-augmented structures--and provide the blueprint for lower layers to execute in real time. By separating plan creation from execution, the DE-Layer enables the use of globally optimized strategies while still allowing adaptive policies at runtime. Requests may arrive in bulk to be jointly optimized, or individually to be served immediately or accumulated for later batching. The DE-Layer also integrates pre-distribution plans into its framework, treating them as background tasks whose outputs are stored in the \get for future use. 

\para{Handling Requests in a Batch.}
The QN stack supports the following strategies for generating entanglement distribution plans for batched requests. A key distinction is whether to iterate over optimized choices for each request in the batch, or to compute a globally optimal structure that jointly serves the entire batch.

\softpara{Iterative Creation of (Purification-Augmented) Swapping Trees.}
To serve a batch of requests, the stack iteratively computes the swapping tree for each request. One potential approach is to iteratively apply an optimal dynamic programming-based method~\cite{swapping-tqe-22}, which constructs an optimal swapping tree for a given EP with minimum expected generation latency by recursively defining and computing appropriate subproblems efficiently. Thus, we process a batch of EP requests by iteratively constructing the optimal swapping tree for each request. After creating a swapping tree for a request, the request's required resources are marked as unavailable to the remaining requests, thereby preventing conflicts when iteratively constructing trees. The dynamic-programming approach has recently been extended to include purification operations~\cite{qcnc-purification} by constructing a purification-augmented swapping tree for each request.


\softpara{Level-Based (Purification-Augmented) Swapping Structure.}
Alternatively, one can directly compute the globally optimal solution for the batch of requests. One approach is to use a linear programming (LP)- based method, which first constructs a hypergraph that encodes all possible level-based swapping structures. In the hypergraph, (i) each hypervertex is a potential intermediate entanglement state, (ii) each hyperedge ($\{s_1, s_2\}, s_3$), for vertices $s_1, s_2, s_3$, is a swapping operation that swaps entanglement states $s_1$ and $s_2$ to create $s_3$,
(iii) a ``hyperpath'' is a potential swapping tree, (iv) and a hyperflow is a level-based structure; here, a hyperflow is a ``combination'' of hyperpaths similar to a network flow in a simple graph being a ``combination'' of simple paths. 
Now, the problem of determining the optimal (maximum aggregate generation rate) level-based structure for a given set of entanglement requests can be formulated as a linear program by assigning flow variables to hyperedges (representing generation rates), with linear constraints that capture network resource limitations and enforce flow conservation. Solving this LP yields the optimal hyperflow and, consequently, an optimal level-based swapping structure~\cite{swapping-tqe-22,qcnc-purification}. The LP approach has also been extended to include purification operations~\cite{qcnc-purification}. 

\softpara{Iterative vs.\ Level-Based Structure.}
When handling entanglement requests in a batch, the stack can choose between the two strategies mentioned above, based on the following trade-offs. The first approach constructs swapping trees for each request individually, allowing flexibility and per-request granularity, but may be suboptimal globally. The second approach computes a globally optimal level-based structure, which can achieve higher aggregate performance but is computationally more expensive~\cite{swapping-tqe-22}. We note that the plan-creation step by the {\tt DE-Layer} is performed offline; hence, some computational overhead is tolerable.

\para{Handling a Single Request.}
An individual entanglement request can be processed immediately or deferred briefly for potential batching. To process a request immediately, there are two options: (a) Without disrupting (or deteriorating the performance of) ongoing plans, the system utilizes the remaining network resources to compute an optimal swapping tree for the new request, using the iterative approach. (b) Alternatively, all ongoing generation plans are suspended, and the latest request is combined with the existing ones into a batch. The system then recomputes the solution for the entire batch using the aforementioned batched request handling methods. The overall strategy should be to use the first approach most of the time, while using the second approach often enough to ensure efficient request generation.

\para{Generating Pre-Distributed EPs.} 
Upon receiving the plans for pre-distributed EPs from the Transport Layer, the Distributed Entanglements Layer translates this specification into swapping trees/structures.
These plans are assigned request identifiers with a reserved value, i.e., \texttt{request\_id = pre\_distributed}, allowing the stack to recognize them and assign a lower scheduling priority. This ensures that they utilize only idle or surplus resources and do not interfere with other running requests. Once the pre-distributed entanglement pairs are successfully generated, they are stored in the \get and become available to serve other future requests via entanglement swapping or purification. 
In the Swapping/Fusion Layer, EPs with \texttt{request\_id = pre\_distributed} are treated as flexible resources whose \emph{request\_id} can be interpreted as matching any active request's ID.

We support two models for generating pre-distributed EPs.
In the \emph{one-time generation model}, a fixed number of EPs is generated before all requests, and no further generation is triggered once the batch is complete.
In contrast, the \emph{continuous replenishment model} monitors the number of pre-distributed EPs for each selected node-pairs and proactively generates additional pairs whenever the available stock falls below a predefined threshold.

\softpara{Role of \get.} 
The \get plays a critical role in enabling pre-distribution by treating these pairs as cached resources that can be dynamically re-labeled to satisfy new requests, thereby bridging offline planning with runtime flexibility.




\subsection{\bf Global Entanglement Module: Support for Adaptive Entanglement Generation}

The \emph{Global Entanglement Module (\get)} is the enabler of adaptive entanglement generation. 
It provides each node with a near-real-time, best-effort view of all currently available entangled pairs, 
together with their key metadata such as estimated fidelity, age, creation time, and assigned request. 

During execution, the Swapping/Fusion Layer queries the local \get replica to select which entanglement pairs should be swapped, purified, or discarded. 
This shared abstraction ensures that local decisions are not made in isolation, 
but are instead aligned with the network's evolving state. 
In particular, adaptive policies such as \emph{youngest}, \emph{oldest}, or the 
\emph{scoring-based strategy} rely on \get metadata to prioritize high-value pairs 
and avoid wasteful operations on degraded ones.

Synchronization across nodes is achieved through lightweight broadcast updates, 
which maintain eventual consistency among local \get replicas. This mechanism is 
sufficient for near-term networks where classical channels are reliable, and the 
scale is modest (e.g., \scyqnet). By tolerating small, temporary inconsistencies, 
the design balances update overhead with decision quality, ensuring that adaptive 
execution remains both scalable and efficient.

Overall, \get transforms adaptive execution from a local heuristic into a 
network-wide coordinated process, providing the foundation on which all 
runtime entanglement generation strategies operate.



\subsection{\bf Swapping Layer: Real-Time Plan Execution}

The previous layer generates static generation plans, which the Swapping Layer then executes within the network, dynamically adapting to real-time variations in entanglement availability due to stochastic generation and decoherence. 

\softpara{Motivation for Adaptive Real-Time Execution.}
We note that the swapping trees created by the previous layer are based on the expected success of swapping and purification operations, and thus, only guarantee {\em expected} performance. However, in practice, one "branch" of a swapping tree may complete far earlier than the other, forcing unexpectedly early EPs to wait in memory longer than anticipated. During this wait, EP fidelity may degrade, or the EP may even become unusable, reducing overall efficiency and success rate. Adaptive execution seeks to minimize such mismatches. Below, we begin by discussing our adaptive execution approach for a single swapping tree and then extend it to broader contexts (e.g., with purification).

\para{Adaptive Swapping over a Single Swapping Tree.}
At runtime, the Swapping Layer essentially adapts a swapping tree (created by the  {\tt DE-Layer}) by determining the best ''swapping order'' (based on the current network state and entanglement availability in \get)---while still keeping the entanglement route unchanged.\footnote{More sophisticated strategies that allow real-time adaptation of the {\em route} also---are deferred to future work.} 
The tree structure may evolve dynamically at runtime, while the default swapping tree provided by the DE-Layer serves both as a fallback configuration—used when no adaptive swapping decision is feasible—and as a baseline for guiding link-EP generation.

To determine the swapping order at runtime, the Swapping Agent at each node consults the \get, which provides a consistent, cross-layer view of entanglement availability. This shared state enables decentralized yet coordinated decisions, ensuring that local swaps align with the global execution context. 
Decisions are primarily based on the age, accumulated latency, and expected regeneration latency of the EPs currently available on the entanglement route. 
Since these determinations must be made in real time, the decision process is designed to be computationally lightweight. With that in mind, we have explored the following adaptive policies.
\begin{enumerate}
\item
\emph{Oldest-First:} Reduces the risk of decoherence by prioritizing EPs with greater age.
\item
\emph{Youngest-First:} Favors fresher EPs to maximize fidelity.
\item
\emph{Longest-Hop:} Prioritizes EPs that have traversed more hops, aiming to accelerate progress on longer entanglement paths.
\item
\emph{Shortest-Hop:} Quickly completes shorter-distance EPs to reduce system-wide congestion.
\item
\emph{Score-Based:} Determines next swapping operation based on a score (discussed below). 
\end{enumerate}
These policies are modular and can be selected or combined based on network conditions or application requirements. The trade-offs between the above policies are as follows. The \emph{Oldest-First} policy proactively acts on the most vulnerable EPs to reduce the risk of being discarded due to decoherence. However, this may delay the use of fresher, higher-fidelity EPs. In contrast, the \emph{Youngest-First} policy prioritizes high-fidelity EPs to maximize output quality. The \emph{Longest-Hop} policy focuses on ``consuming'' EPs that have already incurred high generation latency before they degrade. Conversely, the \emph{Shortest-Hop} policy prioritizes EPs with fewer hops, thereby releasing memory more quickly. 

\begin{algorithm}[t]
\caption{\blue{Swap Scoring Function}}
\label{alg:swap-scoring}

\textbf{Input:}
EP pair $(i, m)$ and $(m, j)$ held at node $m$ for request $r$;
path $P_r = (v_0, v_1, \ldots, v_L)$ with $L$ hops;
GEM replica $\mathcal{E}_{\text{get}}$ at $m$: set of active EPs, each 
with endpoints, request ID, and age; offline plans $\mathcal{R}$,
cutoff age $T_{\max}$;
weighting parameters $\alpha, \beta, \gamma$.

\textbf{Output:}
Swap priority score $S$ for producing EP $(i, j)$.

\begin{algorithmic}[1]

\STATE \textbf{// Step 1: Time urgency}
\STATE $a_{\text{new}} \gets \max\bigl(\text{age}(i,m),\; \text{age}(m,j)\bigr)$
\STATE $U_{\text{time}} \gets \dfrac{1}{T_{\max} - a_{\text{new}}}$
\hfill \COMMENT{Higher when EP is closer to expiry}

\medskip

\STATE \textbf{// Step 2: Global urgency (route completion ratio)}
\STATE Let $\mathcal{E}_r \subseteq \mathcal{E}_{\text{get}}$ be all EPs belonging to request $r$ that are \emph{visible} to node $m$:
\begin{align*}
\mathcal{E}_r = \{(u,v) \in \mathcal{E}_{\text{get}} \mid \mathrm{req}(u,v) = r\}
\end{align*}
\STATE For each EP $(u,v) \in \mathcal{E}_r$, let $\texttt{subroute}(u,v; r)$ be the subset of links in route $r$ covered by $(u,v)$.
\STATE $D_{ij} \gets \dfrac{\bigl|\, \bigcup_{(u,v) \in \mathcal{E}_r} \texttt{subroute}(u,v; r) \,\bigr|}{L}$
\hfill \COMMENT{Route completion ratio $\texttt{rcomp}(r, \mathcal{E}_{\text{get}})$}

\medskip

\STATE \textbf{// Step 3: Overall Urgency Score}
\STATE $\text{U} \gets U_{\text{time}} \times D_{ij}$

\medskip

\STATE \textbf{// Step 4: Swap Progress}
\STATE Let $h_1$ = number of hops spanned by EP $(i,m)$, \; $h_2$ = number of hops spanned by EP $(m,j)$
\STATE $\text{PG} \gets \dfrac{h_1 + h_2}{L}$
\hfill \COMMENT{Fraction of path this swap covers}

\medskip

\STATE \textbf{// Step 5: Opportunity Loss}
\STATE Let $\mathcal{P}$ be the set of all planned swap rules from offline routes $\mathcal{R}$.
\IF{$\exists\, p \in \mathcal{P}$ such that $p$ is exactly the swap of $(i,m)$ and $(m,j)$}
    \STATE $\text{OL} \gets 0$ \hfill \COMMENT{This swap is planned; no lost opportunity}
\ELSE
    \STATE $\text{OL} \gets \sum_{p \in \mathcal{P}}\! \lambda_p \cdot \bigl[\, (i,m) \in p \;\text{or}\; (m,j) \in p \,\bigr]$
    \hfill \COMMENT{$\lambda_p$: generation rate of rule $p$}
\ENDIF

\medskip

\STATE \textbf{// Step 6: Final Score}
\STATE $S \gets \alpha \cdot \text{U} + \beta \cdot \text{PG} - \gamma \cdot \text{OL}$

\medskip
\RETURN $S$
\end{algorithmic}
\end{algorithm}
\softpara{Score-Based Swapping.}
\green{In the score-based adaptive strategy, each node evaluates its locally stored EPs and assigns a score to each eligible swap candidate. For a candidate swap between EPs $(i,m)$ and $(m,j)$ at node $m$, the score consists of three components (Algorithm~\ref{alg:swap-scoring}).}
\begin{packedenumerate}
\item 
\emph{Urgency Score} (U), which combines time urgency and global urgency. {\em Time urgency} is defined as the reciprocal of the remaining lifetime of the resulting EP; the closer the EP is to expiry, the more urgent it is to act. {\em Global urgency} is measured by the route-completion ratio, i.e., the fraction of the target path that is already covered by EPs currently available and visible to node $m$ via \get.
\item 
\emph{Swap Progress} (PG), defined as the fraction of the total path spanned by the two EPs being swapped. A swap that covers more hops in a single operation is preferred, as it advances the distribution of entanglement more efficiently.
\item 
\emph{Opportunity Loss} (OL), which penalizes swaps that deviate from the offline plan. If the candidate swap matches a planned swap rule, no opportunity is lost, and OL is zero. Otherwise, OL is proportional to the expected generation rates of the planned swaps that would be blocked if these EPs were consumed.
\end{packedenumerate}

\green{The final score is computed as $S = \alpha \cdot U + \beta \cdot \text{PG} - \gamma \cdot \text{OL}$. The weighting parameters $\alpha$, $\beta$, and $\gamma$ are determined by coordinate descent over a set of training network instances: starting from an initial estimate, we iteratively optimize one parameter at a time while keeping the others fixed, selecting the value that maximizes the average entanglement generation rate over the training set, and repeating until convergence.}

\green{The node then selects the candidate swap with the highest positive score. If no positive-score swap exists, the node waits until a new local EP is generated and then reevaluates the candidate swaps.}

\para{Comparison with Prior Adaptive Approaches.} Prior adaptive approaches, such as the MDP-based optimization in~\cite{mdp} and the DP-based strategy in~\cite{real-time}, have aimed for theoretical optimality. However, these methods are computationally expensive and thus impractical for real-time decision-making in quantum networks, where entanglement availability fluctuates rapidly. In particular, the MDP-based approach in~\cite{mdp} incurs significantly higher computational overhead because it must explicitly model and solve a large state-action space.
Moreover, these schemes operate under strong assumptions that limit their applicability and generalizability. For instance, both approaches assume that each node has only two memory positions. Additionally, the approach in~\cite{real-time} assumes that once a link-level EP is established, the link becomes inactive and cannot be reused until the EP, along with all its descendants generated through entanglement swapping, is either consumed or lost due to decoherence or failure.

\para{Extensions to Broader Contexts.}

\softpara{Multiple Trees or Level-based Structures.}
The above adaptive execution strategy can also be applied to multiple concurrently executing swapping trees or level-based swapping structures. In the case of a level-based swapping-structure solution for batched requests, one approach is to first decompose the problem into multiple swapping trees. Each tree is then executed independently using the adaptive single-path mechanism described above.

\softpara{Purification-Augmented Plans.}
Beyond adaptive swapping strategies, our protocol stack is also designed to accommodate real-time purification decisions. 
The key enabler here is the \get, which maintains global metadata on fidelity, age, and usage. Without such a shared state, purification decisions would be purely local and risk wasting resources; with the \get, they become globally informed and adaptive, enabling nodes to determine at runtime whether purification will improve overall  rate and fidelity, or whether EPs should be used directly for swapping. Designing effective strategies that exploit this capability remains an open direction for future research.

\softpara{Proactive Strategy for Discarding Decohered EPs.}
To mitigate the impact of decoherence, the Swapping Layer proactively discards EPs that are unlikely to contribute to high-fidelity outcomes. Instead of waiting until the final target EP fails, the system evaluates \emph{intermediate} EPs during execution and discards those that are too aged to be useful. This avoids wasting fresh EPs in swaps with degraded ones, improving both efficiency and fidelity.
\green{Concretely, each EP is assigned an age cutoff proportional to its 
 expected regeneration latency:}
$
T_{\mathrm{cutoff}}(v) = 2 \cdot L_{\mathrm{expected}}(v),
$
\green{where $v$ is the node in the swapping tree representing the EP, 
and $L_{\mathrm{expected}}(v)$ is its expected generation latency, 
computed recursively during plan construction via 
Eq.~\ref{eqn:EP_swaplatency}.}

\subsection{\bf Link Layer: Managing Link-EP Generation} 

The Link Layer manages entanglement generation across direct links in the quantum network by scheduling link-EP requests through low-level physical operations. Its role is to translate higher-layer requirements into concrete generation attempts, satisfy target rates and fidelities, and resolve contention over shared link resources. In addition, the Link Layer performs purification of link-EPs as instructed by higher layers to improve fidelity before passing them upward in the stack.

\para{Handling Link-EP Requests.}
Entanglement between neighboring nodes can be established using three main architectures:
\begin{itemize}
    \item \emph{Meet-in-the-Middle:} both nodes send photons to a central Bell-state measurement (BSM) station.
    \item \emph{Sender–Receiver:} one node emits photons while the other detects them; however, the longer travel distance for each photon reduces fidelity.
    \item \emph{Memory–Source–Memory:} a central entangled-photon source distributes one photon to each node, which is then stored in local quantum memories.
\end{itemize}
Though differing in hardware placement, all approaches rely on precise timing to align detection or storage windows. The Link Layer coordinates these timings, arbitrates access to shared resources, and interfaces with the Physical Layer to execute the underlying hardware operations.


For each entanglement-generation request, the Distributed Entanglements (DE) Layer issues link-EP requests that specify the desired rate and fidelity. The Link Layer aggregates rate requirements across requests and determines how frequently each link should attempt to generate entanglement. It then relies on the Physical Layer to perform synchronized photon emission or detection with minimal latency. At the node level, the Link Layer multiplexes hardware resources such as excitation attempts across incident links, ensuring global synchronization with neighboring nodes.

Upon a successful heralding of a link-EP, the Link Layer assigns it to one of the active requests in the pool according to a scheduling policy (e.g., weighted round-robin). This assignment enforces fairness and responsiveness across concurrent link-EP requests. When higher layers request purification, the Link Layer invokes local purification steps before returning EPs upward, thereby providing a tunable trade-off between fidelity and throughput.

Overall, the Link Layer does not define new physical mechanisms for entanglement generation, but provides the architectural glue that maps diverse hardware architectures to higher-layer abstractions. By handling synchronization, multiplexing, scheduling, and purification at the link granularity, it exposes a uniform service to the stack while insulating higher layers from hardware-specific complexity.

\section{\bf Other Applications}
\label{sec:other-apps}

Beyond generating bipartite entanglement, the proposed stack serves as a foundation for broader quantum applications. 

\subsection{Multipartite Entanglement Generation}  
Multipartite states such as GHZ and graph states are essential resources for distributed computing, sensing, and cryptography~\cite{ghz-qce-23,fan2024optimized}. Their creation requires coordinated use of entanglement swapping and fusion.  

\begin{itemize}
    \item \emph{Specification:} A GHZ request specifies a set of nodes $S \subseteq V(Q)$, with one qubit of the GHZ state assigned to each. Graph-state requests specify a target graph $G$ and a distribution map $\tau: V(G) \mapsto V(Q)$.  
    \item \emph{Execution:} The Distributed Entanglements Layer plans multipartite generation strategies, extending swapping-tree abstractions with fusion operations.
    Prior work~\cite{korean-graph} established techniques for limited graph-state topologies. In our own work, we developed distribution techniques for GHZ~\cite{ghz-qce-23} and, recently, for arbitrary graph states~\cite{fan2024optimized}.
    \item \emph{Outlook:} Current methods rely on static plans, with no adaptive or pre-distribution strategies yet available. Extending adaptive execution and Global Entanglement Tracking (\get) to multipartite states remains a natural, open direction for future work.  
\end{itemize}

\subsection{Distributed Quantum Computing}  
DQC interconnects multiple quantum processors to execute algorithms beyond the capacity of any single device~\cite{caleffi}. The main challenge is generating entangled pairs (EPs) in time to respect gate dependencies while mitigating decoherence.  

\begin{itemize}
    \item \emph{Application Layer:} DQC workloads are expressed as \emph{Distributed Circuit Entanglement} (DCE) requests, specifying EPs together with their consumption order.  
    \item \emph{Transport Layer:} Computes batching plans that group EPs into temporally aligned sets, ensuring they can be generated within decoherence limits.  
    \item \emph{Execution:} Each batch is scheduled and executed by the Distributed Entanglements Layer, with lower layers carrying out entanglement generation as in the standard process. Failed batches trigger replanning before subsequent batches proceed.  
\end{itemize}

In both settings, the layered stack provides the abstractions and control mechanisms needed to translate high-level requirements into executable entanglement plans, while highlighting open opportunities to extend adaptive and predistribution techniques beyond the bipartite case.




  

\begin{figure*}[t]
\vspace*{-0.3in}
    \centering
    \includegraphics[width=\textwidth]{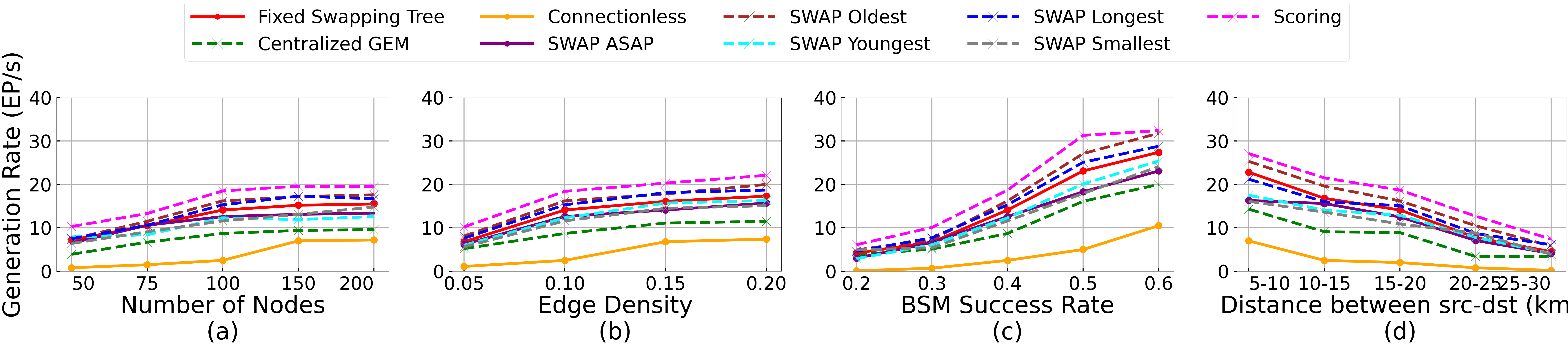}
    \caption{Generation rates of entangled pairs under varying parameters. The \emph{Scoring} strategy consistently achieves the highest rates, outperforming all other adaptive heuristics as well as {\em Fixed-tree} and {\em Connectionless} baselines. Performance of the \emph{centralized module} scheme highlights the cost of centralizing \get, underscoring the necessity of a distributed \get for scalable adaptive execution.}
    \vspace*{-0.1in}
    \label{fig:adaptive-4-parameters}
\end{figure*}

\begin{figure*}[t]
    \centering
    \includegraphics[width=\textwidth]{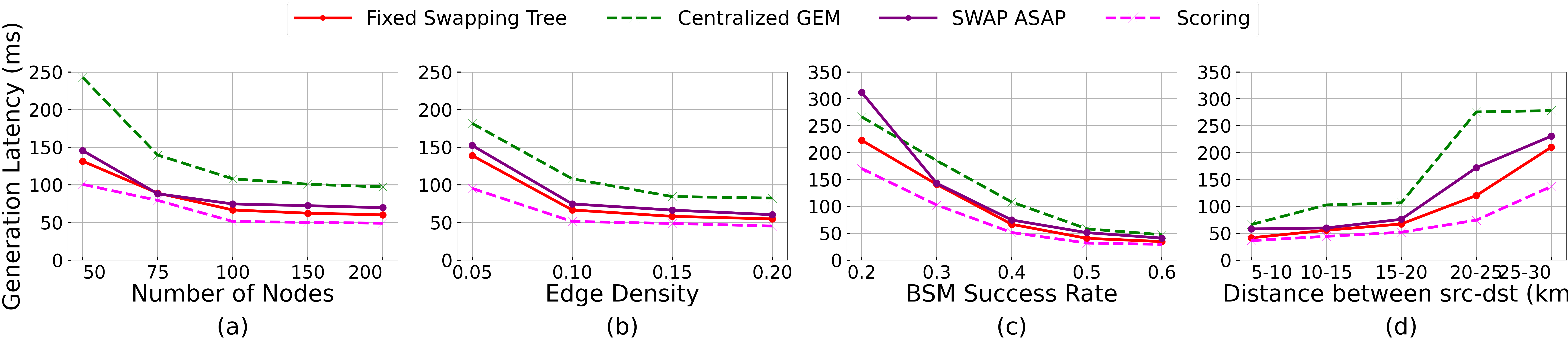}
    \caption{Average generation latency under the one-time batch predistribution model. While predistribution improves performance over no predistribution, benefits taper off as the stock is depleted. Adaptive strategies that use the distributed \get, particularly \emph{Scoring}, extract the most value from these cached entanglement resources.}
    \vspace*{-0.1in}
    \label{fig:predistribution-once-4-parameters}
\end{figure*}

\begin{figure*}[t]
\vspace*{-0.1in}
    \centering
    \includegraphics[width=\textwidth]{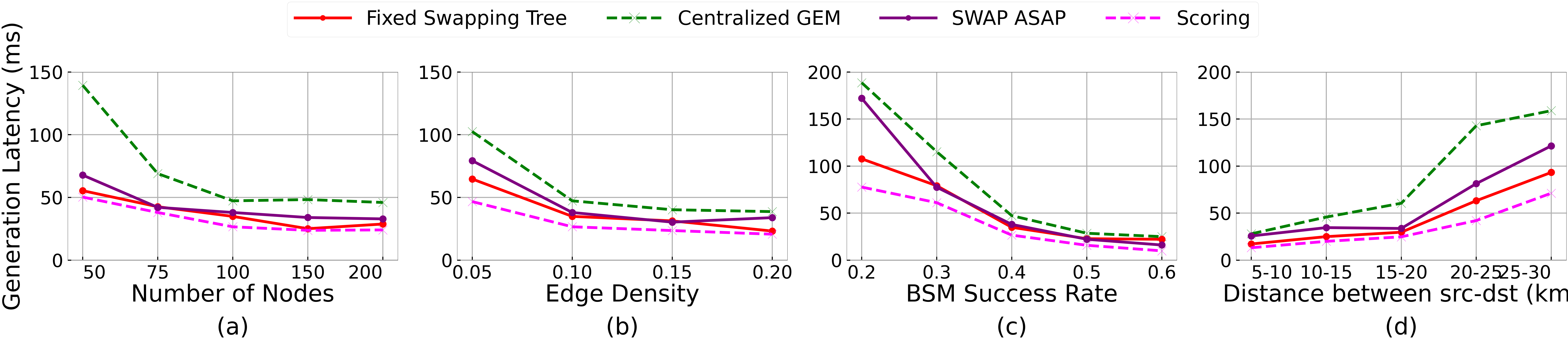}
    \caption{Average generation latency under the continuous predistribution model. Maintaining a replenished stock of predistributed EPs significantly reduces latency compared to one-time prefetching. The \emph{Scoring} strategy again delivers the largest gain, showing that proactive resource caching and adaptive \get-guided execution work synergistically.}
    \vspace*{-0.2in}
    \label{fig:predistribution-continuous-4-parameters}
\end{figure*}

\begin{figure*}[t]
\vspace*{-0in}
    \centering
    \includegraphics[width=0.9\textwidth]{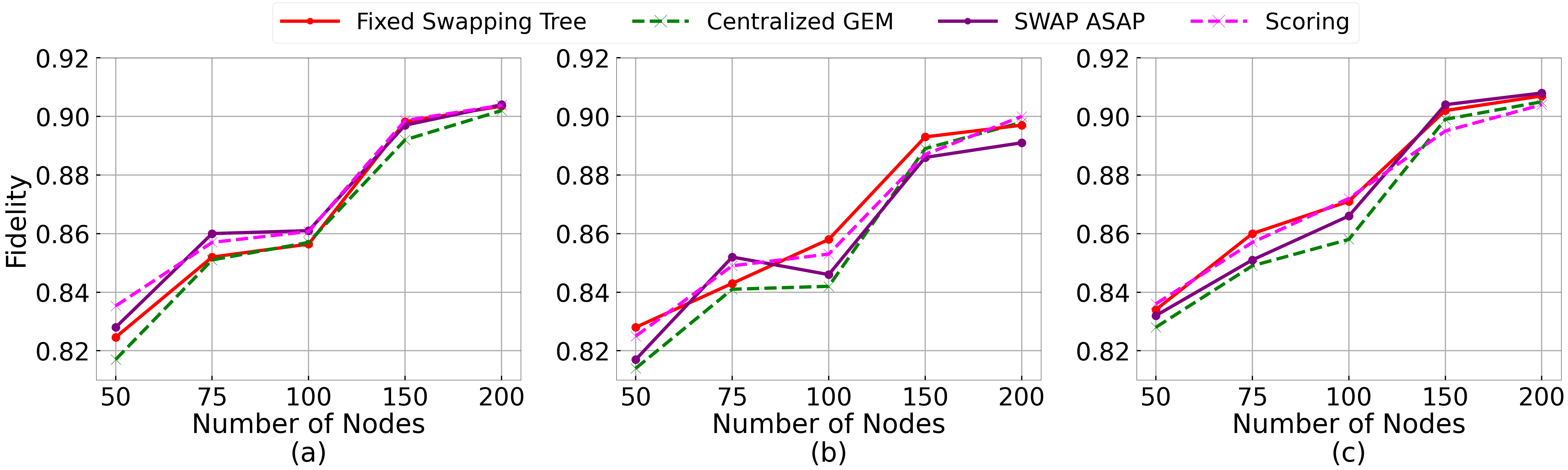}
    \caption{Fidelity under different predistribution settings.
        (a) \emph{No Predistribution}, 
        (b) \emph{One-time Batch Predistribution Model}, 
        (c) \emph{Continuous Predistribution Model}.}
    \vspace*{-0.2in}
    \label{fig:fidelity-nodes}
\end{figure*}

\section{\bf Evaluations}
\label{sec:eval}

\para{NetSquid Protocol.}
We build our protocols on top of the link-layer protocol of~\cite{sigcomm19}, delegated to continuously generate \epss on a link at a desired rate. 
The protocol stack is organized into independent layers, with each layer containing a set of protocols that operate within its own scope, as described earlier in the stack architecture.

\para{Simulation Setting.}
We adopt the standard random quantum network setup used in prior works~\cite{swapping-tqe-22, sigcomm20, fan2024optimized}.
By default, we use a network spread over an area of $100 km \times 100 km$.
We use the Waxman model~\cite{waxman}, used to create Internet topologies,
distribute the nodes, and create links. 
\green{We adopt the meet-in-the-middle architecture for link-EP 
generation: each pair of adjacent nodes generates atom-photon EPs 
locally and transmits photons to a photon-photon BSM 
device at the midpoint~\cite{swapping-tqe-22}.}
We vary the number of nodes from 50 to 200 (default: 100); each node has 5 memory positions; and the edge density ranges from 0.05 to 0.2 (default: 0.1).
Each data point is for a 100-second simulation in NetSquid.

\softpara{Parameter Values.}
We use parameter values similar to the ones used in~\cite{caleffi, swapping-tqe-22}.
In particular, we use swapping probability of success (\fp) to be 0.4 and latency (\ft) to be 10 $\mu$ secs; 
in some plots, we vary $\fp$ from 0.2 to 0.6.
The atomic-BSM probability of success (\bp) and latency (\bt) always equal their fusion counterparts \fp and \ft. 
The optical-BSM probability of success (\php) is half of \bp. 
To generate link-level \epss, we use atom-photon generation times (\gt) and the probability of success (\gp) of 50 $\mu$sec and 0.33, respectively. 
Finally, we use photon transmission success probability 
as $e^{-d/(2L)}$~\cite{caleffi} where $L$ is the channel attenuation length
(chosen as 20km for optical fiber) and $d$ is the distance between the nodes.
The default cut-off age for target EPs at the root of the swapping tree is set to twice the expected generation latency.
For the experiments involving predistribution, we generate $10$ EPs for each selected node pair for both the one-time batch model and the continuous predistribution model.
And there are $15$ different requests, each continuously generating EPs between a distinct source-destination node pair, 
and $5$ node pairs are selected to generate predistributed EPs. 
For both models, the interval between two consecutive EP generation requests is set to $5\,\mathrm{s}$.
Fidelity is modeled in NetSquid using two parameter values: depolarization (for decoherence) and dephasing (for operation-driven) rates; we choose a depolarization rate of 0.01 and a dephasing rate of 1000~\cite{swapping-tqe-22,chakraborty2020entanglement}.

\para{Prior Algorithms Compared.}
For comparison with the adaptive routing strategies newly supported by our protocol stack, we evaluate against the following baselines: (i) the \emph{Fixed-Swapping-Tree} approach from~\cite{swapping-tqe-22}, which uses a dynamic programming approach to create an {\em optimal} swapping tree, and (ii) the \emph{Connectionless} hop-by-hop routing strategy implemented within a protocol stack as in~\cite{towsley-24}. 
Together, these baselines capture the two dominant design philosophies: globally optimal but rigid trees and fully local, connectionless routing.
We do not include other {\em adaptive} routing strategies such as those in~\cite{real-time, mdp}, as they rely on strong assumptions about node memory, that each node has only two memory positions. Moreover, their online decision-making incurs significant computational overhead.

\para{Our Algorithms.}
In our experiments, we evaluate five adaptive execution strategies. \emph{youngest}, \emph{oldest}, \emph{longest-hop}, \emph{shortest-hop}, and \emph{scoring} as described in Section~\ref{sec:Functionality}. 
In all five strategies, nodes maintain local copies of the \get, enabling decentralized yet globally consistent adaptive decisions; each is applied in conjunction with the \emph{Early Discarding Strategy}.
We also implemented a \emph{Centralized Module} scheme, in which we adopt the \emph{oldest} adaptive strategy, but only the node that is physically located at the center of the network stores the \get. All other nodes must send messages to this central node to query the \get when making adaptive online decisions.

For the experiments involving predistribution, we evaluate two models for generating predistributed EPs. The first is the \emph{one-time batch predistribution model}, where a fixed number of EPs are generated in a single batch before request arrivals. The second is the \emph{continuous predistribution model}, in which the system monitors the number of available predistributed EPs and proactively generates additional EPs whenever the stock falls below a predefined threshold.

\para{Evaluation Results.}

\softpara{EP Generation Rate without Predistribution.}
Figs.~\ref{fig:adaptive-4-parameters} report the entanglement generation rates achieved under different adaptive strategies. We vary one parameter at a time while keeping all others fixed at their default values. Among the adaptive policies, the \emph{Scoring} strategy consistently delivers the highest generation rates, outperforming both simple heuristics and prior \emph{fixed-swapping-tree} (by around 20\%) or \emph{connectionless} (by more than 100\%) baselines across all parameter settings. 
\green{As shown in Fig.~\ref{fig:adaptive-4-parameters}(c), the performance-ratio of \emph{Scoring} vs.\ \emph{Connectionless} narrows as $p_f$ increases, since higher BSM success rates reduce the penalty due to cascaded retries in sequential hop-by-hop swapping; } 
The \emph{Oldest} and \emph{Longest} policies also improve over non-adaptive baselines, though less markedly than \emph{Scoring}. In contrast, the \emph{centralized module} scheme performs significantly worse than even the fixed-tree baseline due to the communication latency incurred when every adaptive decision requires querying a single central \get. This result underscores the architectural importance of distributing the \get, which enables decentralized yet globally consistent decisions without bottlenecks.

\softpara{Latency under Two Predistribution Settings.} Figs.~\ref{fig:predistribution-once-4-parameters} and \ref{fig:predistribution-continuous-4-parameters} present results after enhancing the adaptive strategies with predistributed EPs.
For the sake of clarity and presentation, Figs.~\ref{fig:predistribution-once-4-parameters} and \ref{fig:predistribution-continuous-4-parameters} only include the most representative strategies, viz., \emph{Scoring}, \emph{SWAP ASAP}, {\em Centralized} and {\em Fixed Swapping Tree}.
The {\em continuous} predistribution model achieves consistently lower average generation latency than the {\em one-time batch model}, as it maintains a steady supply of entanglement resources in the background. Conceptually, this mirrors a caching system: the one-time model corresponds to static prefetching. In contrast, continuous predistribution corresponds to dynamic replenishment, which ensures that EPs are “on hand” when new requests arrive. In both models, the \emph{Scoring} strategy continues to outperform all alternatives, demonstrating that adaptive \get-guided execution and proactive resource management complement predistribution. 

\softpara{Fidelity under Various Settings.} 
Finally, beyond rate and latency, we evaluate fidelity under the same settings.
In Fig.~\ref{fig:fidelity-nodes}, we observe that fidelity across schemes remains comparable, with the \emph{Scoring} strategy again near the top. Importantly, no adaptive gain in rate or latency comes at the expense of fidelity. We also observe that the end-to-end fidelity increases as the number of nodes grows; this improvement arises because more available nodes provide greater routing flexibility, enabling the selection of shorter paths and thus lower latency.

\begin{figure}[t]
    \centering
    \vspace{-2pt}
    \includegraphics[width=0.9 \columnwidth]{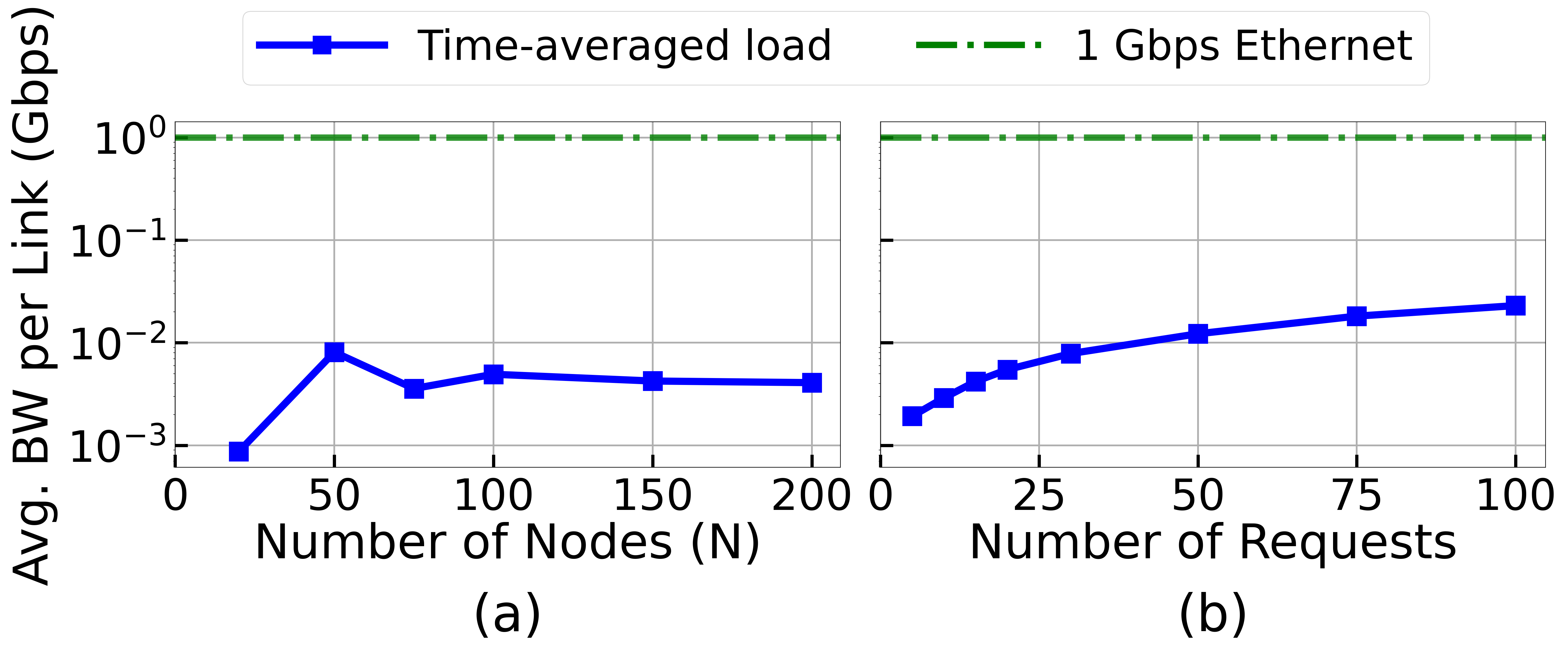}
    \vspace{-1pt}
    \caption{\blue{Classical communication overhead of \texttt{GEM}: per-link bandwidth vs. (a) network size and (b) number of concurrent requests.}} 
    \vspace{-0.2in}
    \label{fig:broadcast-overhead}
\end{figure}

\softpara{Classical Communication Overhead.}
\green{
A natural question is whether the classical control traffic introduced by \texttt{GEM} can be supported by the underlying classical network. In our design, \texttt{GEM} disseminates entanglement-state updates via lightweight broadcast messages that carry EP identifiers, timestamps, hop information, and fidelity metadata. Under our message model, each update packet is approximately 120 bytes, including Ethernet, IP, and UDP headers.
Fig.~\ref{fig:broadcast-overhead} shows the resulting time-averaged per-link bandwidth as the network size and number of requests vary. In all cases, the measured load peaks at only 10--30 Mbps. Thus, as expected, the broadcast overhead of \texttt{GEM} is negligible relative to typical classical link capacities, which are usually on the order of a few Gbps.}

{\em Latency Overhead.} \green{Because \texttt{GEM} replicas across nodes are synchronized via broadcast updates, there is an inherent delay between the occurrence of an entanglement event and the time at which remote nodes become aware of it. As a result, adaptive decisions may occasionally be made using slightly stale information. To capture this effect, we explicitly model classical communication latency in the message-delivery process of our simulations, and all reported results include its impact. Our results show that adaptive execution remains effective despite these broadcast delays.}

\softpara{Summary.} Together, these results validate the central premise of our stack: that the \get enables lightweight, distributed adaptivity which, when combined with proactive resource predistribution, significantly improves efficiency and robustness.


               
\section{\bf Conclusion}  
\label{sec:conc}

We have presented a quantum network protocol stack that integrates a Global Entanglement Module (\get) to enable real-time adaptive entanglement generation, routing, and purification. The design unifies offline planning, runtime adaptive execution, pre-distributed EP utilization, and multipartite state generation within a coherent architecture. Our evaluation results further support the practicality of this design.
%
The architecture is also well-positioned for experimental prototyping. In particular, it is being considered for deployment within \scyqnet, one of the inaugural NSF NQVL projects~\cite{scyqnet} spanning Stony Brook University, Brookhaven National Laboratory, Columbia University, and Yale University. With approximately ten nodes and several quantum processors, \scyqnet provides a realistic near-term platform for validating scaled implementations of the proposed stack.

Our results establish a foundation for scalable quantum communications. While \get is viable for near-term networks such as \scyqnet, scaling it to larger deployments will require approximate or hierarchical synchronization mechanisms. More broadly, the introduction of \get as a dedicated cross-layer coordination module distinguishes our architecture from prior designs and paves the way for robust, adaptive quantum internet systems.

\bibliographystyle{IEEEtran}
\bibliography{all_shorten}

\end{document}